# Driving non-trivial quantum phases in conventional semiconductors with intense excitonic fields

Vivek Pareek[1†], David R. Bacon[1†], Xing Zhu[1†], Yang-Hao Chan[2], Fabio Bussolotti[3], Nicholas S. Chan[1], Joel Pérez Urquizo[1], Kenji Watanabe[4], Takashi Taniguchi[5], Michael K. L. Man[1], Julien Madéo[1], Diana Y. Qiu[6], Kuan Eng Johnson Goh[3,7,8], Felipe H. da Jornada[9, 10*], Keshav M. Dani[1*]

[1] Femtosecond Spectroscopy Unit, Okinawa Institute of Science and Technology Graduate University; Onna, Okinawa, Japan 904-0495

[2] Institute of Atomic and Molecular Sciences, Academia Sinica, and Physics Division, National Center of Theoretical Sciences; Taipei, Taiwan

[3] Institute of Materials Research and Engineering (IMRE), Agency for Science, Technology and Research (A*STAR); 2 Fusionopolis Way, Singapore, 138634 Singapore

[4] Research Center for Functional Materials, National Institute for Materials Science; 1-1 Namiki, Tsukuba, Ibaraki 305-0044, Japan

[5] International Center for Materials Nanoarchitectonics, National Institute for Materials Science; 1-1 Namiki, Tsukuba, Ibaraki 305-0044, Japan

[6] Department of Mechanical Engineering and Materials Science, Yale University; New Haven, CT, USA

[7] Department of Physics, National University of Singapore; 2 Science Drive 3, Singapore, 117551 Singapore

[8] Division of Physics and Applied Physics, School of Physical and Mathematical Sciences, Nanyang Technological University; 50 Nanyang Avenue, Singapore 639798, Singapore

[9] Department of Materials Science and Engineering, Stanford University; Stanford, CA, USA

[10] Stanford PULSE Institute, SLAC National Accelerator Laboratory; Menlo Park, CA, USA

† These authors contributed equally to this work, * Corresponding authors.

# Abstract

Inducing novel quantum phases and topologies in materials using intense light fields is a key objective of modern condensed matter physics, but nonetheless faces significant experimental challenges. Alternately, theory predicts that in the dense limit, excitons – collective excitations composed of Coulomb-bound electron-hole pairs – could also drive exotic quantum phenomena. However, the direct observation of these phenomena requires the resolution of electronic structure in momentum space in the presence of excitons, which became possible only recently. Here, using time- and angle-resolved photoemission spectroscopy of an atomically thin semiconductor in the presence of a high-density of resonantly and coherently photoexcited excitons, we observe the Bardeen-Cooper-Schrieffer (BCS) excitonic state – analogous to the Cooper pairs of superconductivity. We see the valence band transform from a conventional paraboloid into a Mexican-hat like Bogoliubov dispersion – a hallmark of the excitonic insulator phase; and we observe the recently predicted giant exciton-driven Floquet effects. Our work realizes the promise that intense bosonic fields, other than photons, can also drive novel quantum phenomena and phases in materials.

# Main text

## Introduction

In semiconductors, a key manifestation of the long-range Coulomb interaction is the formation of few-particle bound states, such as excitons and trions, which in turn exhibit novel quantum properties and phases with varying particle density. For example, in the low-density or dilute exciton limit, where inter-excitonic distances are much larger than their size, excitons act as tightly bound, neutral, composite bosons[1]. This limit, extensively studied experimentally and well-described theoretically by the Bethe-Salpeter equation (BSE), displays a distinct optical absorption peak in the forbidden bandgap region of the semiconductor and insulating behavior in transport measurements[2–4]. With coherence, whether imbibed from the photoexciting laser field[5] or from spontaneous condensation at low temperatures[6], the dilute exciton gas forms the Bose-Einstein condensate (BEC) phase. At the other end of the density scale, where the expected inter-excitonic distance is much smaller than the exciton size, one resides well beyond the so-called Mott transition[7–9]. In this very high-density regime, the Coulomb interaction between electrons and holes is strongly screened, and excitons dissociate into unbound electrons and holes. Consistent with experiments, in this regime, the excitonic absorption peak[7,9,10] and intra-excitonic Rydberg transitions are no longer observable[4,11]; the system exhibits metallic properties[8,12], and one observes a giant renormalization of the bandgap[7,13] and core-levels[14], consistent with the presence of a large number of free electrons and holes.

Between these two extreme regimes, at densities substantially higher than the dilute limit, but still short of the Mott transition, striking quantum phenomena have been predicted, from decades ago[1,15], till even recently[16,17]. In this intermediate regime, where the inter-excitonic distance is comparable to the exciton size, the pair-wise Coulomb interactions between electrons and holes continue to play an important role, as indicated by the presence of a clear excitonic absorption peak. Nonetheless, excitons are no longer well-described perturbatively by the BSE as a single correlated electron-hole pair on top of an insulating ground state. Instead, the Coulomb-bound electron hole pair can be thought of analogously to the phonon-bound Cooper pairs of the BCS theory of superconductivity[1,5,15]. Analogous to the physics of excitonic insulators (XI)[18,19] and strongly photoexcited non-equilibrium excitonic insulators (NEQ-XI)[16], the presence of a high-density of excitons alters the quasiparticle self-energy of the material. The quantum nature of the excitonic wavefunction, which can be expressed as a superposition of electron-hole pairs at different wavevectors, transforms the quasiparticle dispersion from a conventional paraboloid to a Mexican-hat-like Bogoliubov dispersion. Such a dispersion change is in stark contrast to the relatively straightforward redshift of the bandgap in the presence of free carriers[7–9,13]. Recently, it has also been proposed that excitons in the dense limit can induce non-trivial topologies and giant Floquet effects in conventional semiconductors[17,20], offering an alternate route to proposed light-driven Floquet schemes, that face challenges related to unwanted absorption, sample heating and damage [21].

While most experiments studying the density dependent phases of excitons have observed the clear and contrasting signatures of the two extreme density regimes, at intermediate densities they have been limited to seeing transitionary effects, such as the gradual broadening and diminishing of the exciton absorption peak. A few optical, terahertz and photoemission spectroscopy experiments have explicitly studied the intermediate dense-exciton regime[5,9,22], but have generally been able to report only the energy shifts at the band-edge, due to the lack of suitable energy *and* momentum resolution simultaneously. The exciton-induced electronic structure changes, such as the features of the BCS phase and the Bogoliubov transformation of the valence band, require measurements over larger regions of momentum space and have yet remained unobservable.

Here, by performing time- and angle-resolved photoemission spectroscopy of monolayer (1L) $WS_2$ in the presence of resonantly and coherently photoexcited excitons of varying densities, we observe the exciton-induced electronic structure changes over the entire Brillouin zone (BZ). As we transition from the dilute-exciton regime to the dense-exciton regime, we observe the predicted features of the BEC to BCS transition of excitons. Concurrently, we observe the transformation in the valence band dispersion – going from the trivial downward curving paraboloid, to the Mexican-hat-like Bogoliubov dispersion. A rigorous, ab initio TD-aGW calculation supports our experimental observations[17]. Finally, in accordance with recent theoretical proposals, we explain how these experimental results can also be understood in the context of exciton-driven Floquet phenomena, confirming orders of magnitude stronger exciton-driven Floquet coupling in practice compared to optically driven Floquet effects.

### ARPES on unperturbed 1L $WS_2$

For our study, we used a CVD-grown 1L $WS_2$ – a prototypical atomically thin semiconductor – transferred onto a low-resistivity Si wafer with a few layers of hBN buffer in between, as in previous studies[23–25]. The sample was placed in a time of flight momentum microscope (ToF

MM)[26,27] at sub 100 K, and a small pristine area of ~16μm x 16μm was selected (Fig. 1b), from which a three-dimensional photoemission dataset was obtained, simultaneously resolving the energy (E) and the two momentum axes ($k_x$, $k_y$)[23–25,28–31] (Fig 1b and supporting material for details). A standard two-dimensional representation of the dataset along the Γ-K-M direction (Fig. 1c) shows the salient features of the 1L WS$_2$ bandstructure: valence band maxima (VBM) at the K points, with a 450 meV spin splitting between the highest (VB1) and second (VB2) valence bands. Improvements in energy resolution of the setup and sample quality yield a FWHM linewidth of around 60meV for VB1[32] (Fig. 1c). Conventional optical absorption spectroscopy on the sample identifies the A exciton resonance peak at 2.1 eV (Fig 1d).

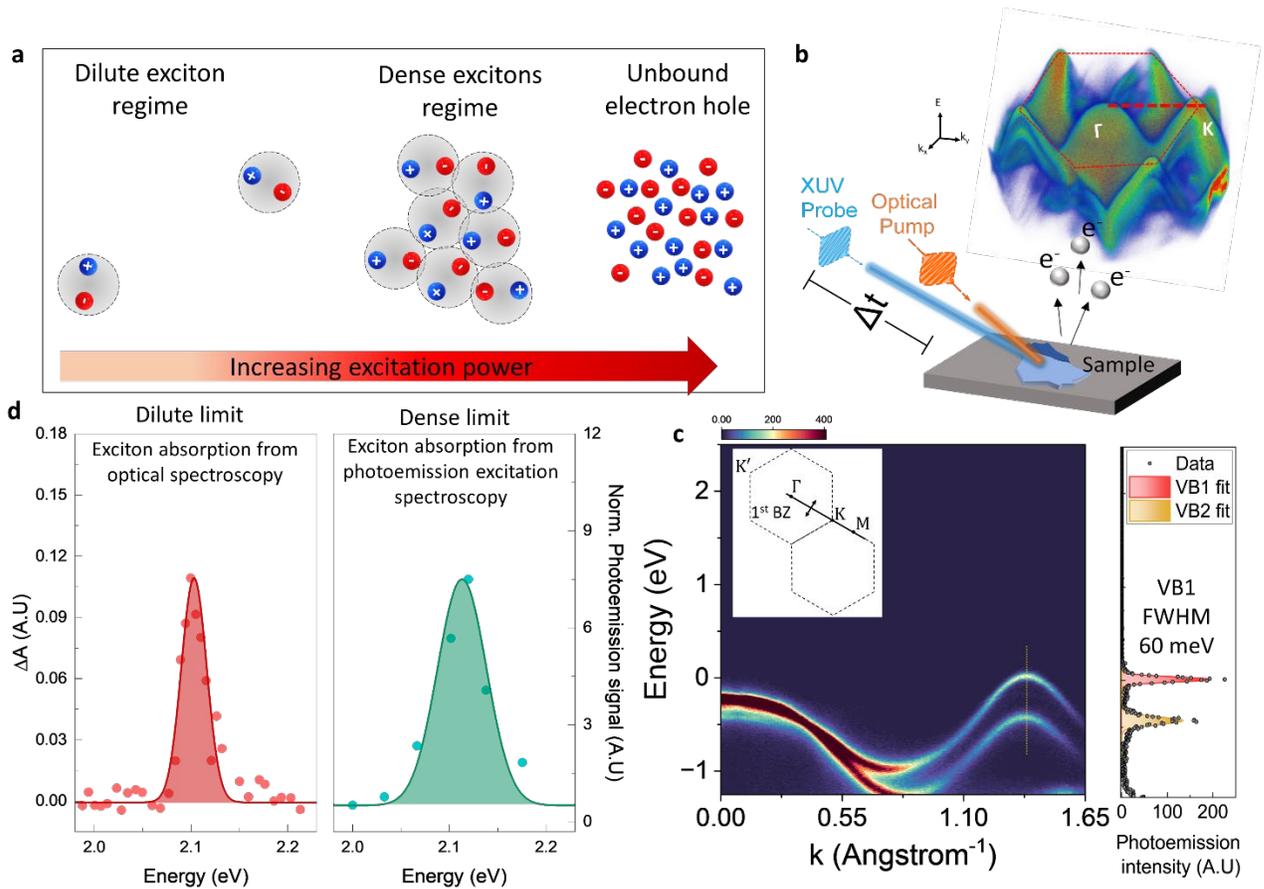

**Figure 1. Dilute to dense exciton regime and experimental configuration.** (**a**) A schematic of the photoexcited exciton density regime with increasing excitation power. In the dilute exciton regime, the inter exciton distance is much larger than the exciton size, whereas in the dense exciton limit the inter exciton distance becomes comparable to the exciton size. Further increasing the exciton density results in the exciton Mott transition such that the excitons dissociate into unbound electrons and holes. (**b**) The experimental schematic of the TR-ARPES and the three-dimensional bandstructure (E, $k_x$, $k_y$) of monolayer WS$_2$ over the entire first Brillouin zone (**c**) Energy-momentum cut along the Γ-K-M direction with the energy distribution curve plotted at the K point. (**d**) Exciton absorption peak in the dilute limit obtained by the optical absorption spectroscopy and the exciton resonance peak in the dense limit obtained by the photoemission excitation spectroscopy. The solid lines are gaussian fits to the data.

**Photoexcited excitons in 1L WS2: Dilute to the dense exciton regime**

Next, to observe the impact of the excitons on the electronic structure of the system, we resonantly excite the A excitons using a 2.1 eV, 100 fs pump pulse and measure the TR-ARPES signal during optical excitation (i.e., around zero pump-probe delay). As reported previously, under these conditions, the photoexcited excitonic population imbibes the coherence from the optical pump field [5].

For low photoexcitation intensities, i.e., in the dilute exciton limit ($n_X = 4 \times 10^{11}$ cm$^{-2}$), we observe the previously reported negative dispersion[24] from the excitonic state appearing 2.1 eV above the valence band (Fig. 2a). In this limit, consistent with the perturbative picture, we do not observe any significant change in the valence band dispersion compared to the equilibrium bandstructure. Previously, theoretical studies of NEQ-XI have described this momentum-resolved photoemission feature from a resonantly and coherently excited, low-density population of excitons as characteristic of the BEC excitonic phase[16].

Now, to explore the dense-exciton regime, we begin to gradually increase photoexcitation intensity. We concurrently monitor the absorption spectrum via a photoemission excitation experiment[23,24] (see also methods) and thereby ensure that we remain in the intermediate dense-exciton regime via the clear presence of the excitonic absorption peak (Fig. 1d). The high-density of coherent excitons creates a non-zero order parameter $\Delta_k = \langle c_{ck} c_{vk}^\dagger \rangle$ in the system, where $c_{n\mathbf{k}}$ is a destruction operator for an electron at band $n$ and wavevector $\mathbf{k}$, and $c$ and $v$ label conduction and valence bands, respectively. This drives a change in the quasiparticle self-energy ($\delta \Sigma^X$) proportional to the order parameter[16,17]. The nature of the excitonic wavefunction $|X\rangle = \sum_{cv\mathbf{k}} A_{cv\mathbf{k}} c_{c\mathbf{k}}^\dagger c_{v\mathbf{k}} |0\rangle$, with an envelope function $A_{cv\mathbf{k}}$ that is strongly peaked at the K and K' points, provides a strong $k$-dependence to the order parameter, and correspondingly to the change in quasiparticle self-energy, which is directly observed in our experiment as a $k$-dependent transformation of the valence band dispersion and the excitonic dispersion at 2.1eV (Fig 2a-d). Accordingly, in the experiment, we observe the two dispersions begin to change around $n_X = 1 \times 10^{12}$ cm$^{-2}$ (Fig 2 b,c). Ultimately, for $n_X = 3 \times 10^{12}$ cm$^{-2}$, they reverse curvature locally at the center of the K valley, where the amplitude of the excitonic envelope function is the largest, driving the largest local change in the quasiparticle self-energy (Fig 2d). The resulting convex-concave band dispersions are a characteristic feature of the BCS phase of excitons[16]. The observed convex-concave *valence band* dispersion is also the hallmark of the quasiparticle dispersion of an equilibrium XI[18,19], where a high density of excitons exists in the ground state itself. In this case, the excitonic dispersion feature (seen at 2.1 eV in our experiment) perfectly overlaps with the valence band dispersion for a direct gap XI.

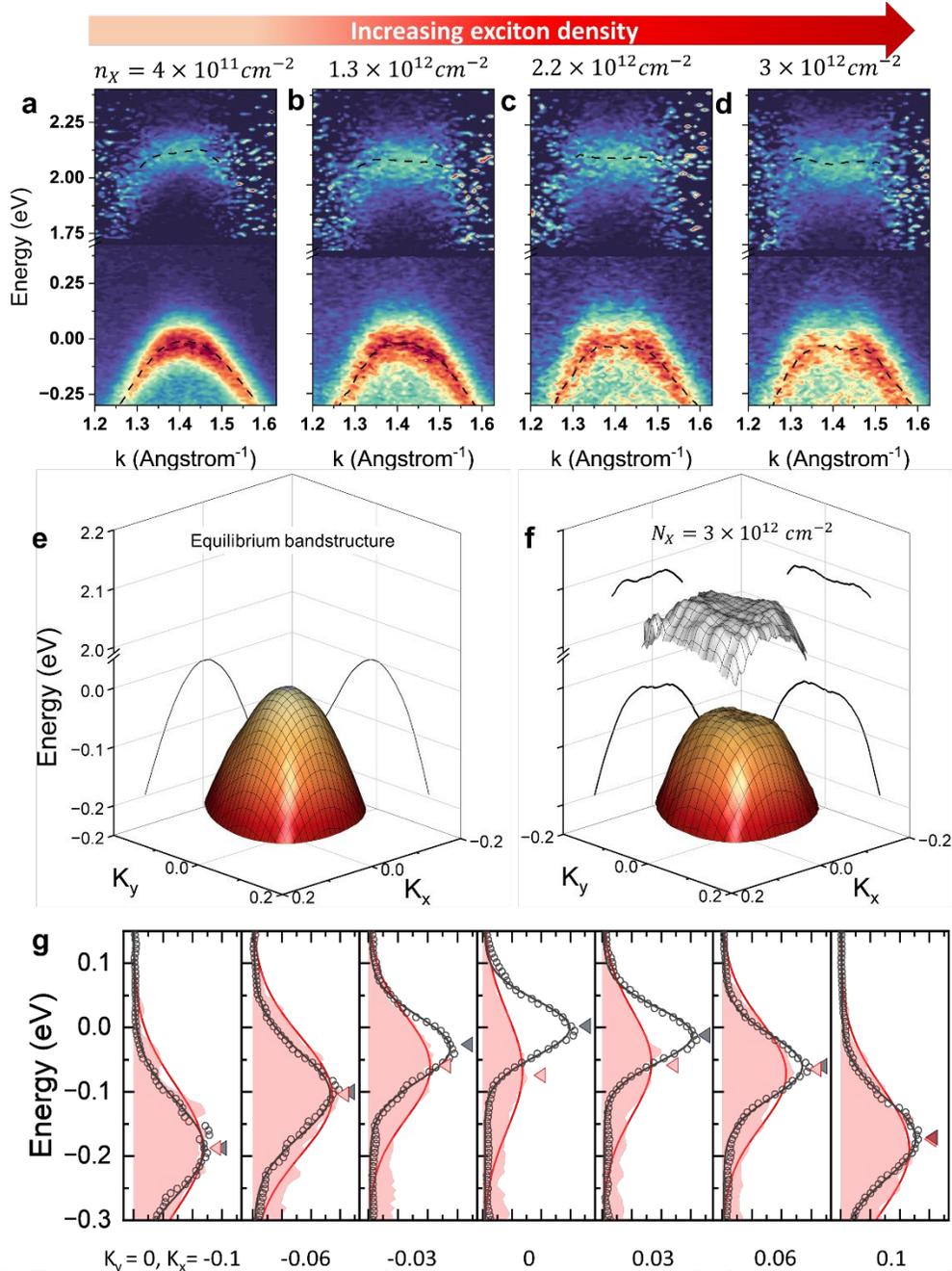

**Figure 2. Effect of increasing exciton density on the quasiparticle bandstructure and exciton-driven Bogoliubov dispersion**: (**a-d**) Energy-momentum ARPES spectra of the valence band and the exciton feature in 1L WS$_2$ with increasing exciton density as measured. The ARPES data was normalized at each k pixel to ensure uniform distribution of the intensity to track the change in the shape of the bandstructure correctly. Experimentally extracted three-dimensional energy-momentum dispersions of the valence band and the exciton bound electron in 1L WS$_2$ around the K point when (**e**) unperturbed (equilibrium), and (**f**) in the dense exciton regime. In (f) we see the distinct Mexican-hat-like Bogoliubov dispersion for both the valence band and the exciton bound electrons indicating a BCS like exciton condensate. Back panel projections show the dispersion relationship along the K$_x$ = 0 and K$_y$ = 0 axes. (**g**) Energy distribution curves (EDC) for different momenta K$_x$ around the K point (K$_y$=0) for the equilibrium band (black circles) and the dense exciton regime (red filled curves). The solid lines are gaussian fits to the EDCs and the peak position extracted from the fits are marked with black (unperturbed) and red (dense exciton regime) triangles.

**Exciton-driven Bogoliubov dispersion**

The convex-concave dispersion plotted in the three-dimensional energy-momentum space resembles a Mexican-hat structure: the K-valley center hosts a local minimum while the VBM occurs along a ring located at a finite momentum away from the K-valley center (Fig 2f). Analogous to the Bogoliubov dispersion of superconductors and of quasiparticles in the XI phase, the VBM ring at finite $k$ represents the breaking of a spontaneous gauge symmetry[33,34], and is expected to support massless (gapless) Nambu-Goldstone excitation modes along the VBM ring, and massive Higgs modes along the radial direction.

To ensure that the observed Mexican-hat-like dispersion is indeed due to a change in the quasiparticle energy, and not an artefact of the exciton-bound holes in the valence band, we compared the measured energy distribution curve at each $k$-vector before photoexcitation in the ground state, and at zero pump-probe delay in the presence of a high-density of excitons. The measurements are carefully taken with the same probe intensity and integration time such that the photoemission signals are directly quantitatively comparable. A few such comparisons are shown in Fig. 2g. In the case of the ground-state, we see a symmetric, Gaussian-shaped energy distribution with a FWHM of ~60 meV at the center of the K-valley (corresponding to the relatively low inhomogeneous broadening of our sample), and a slightly more broadened distribution (FWHM ~60 – 100meV) as we move away from the center of the K-valley (due to the shorter lifetime of holes away from the valley center). For the measurements in the presence of a high-density of excitons, we continue to see a *symmetric, Gaussian-shaped* energy distribution, now with a k-dependent peak-energy shift, broadening and decreased signal (i.e., area under the curve). The $k$-dependent decrease in the signal exactly corresponds to the density of exciton-bound holes at the given $k$-vector, as reported previously[25]. In contrast, the $k$-dependent *shift in peak energy* corresponds to the change in the dispersion of the valence band discussed in this manuscript. As can be seen (Fig. 2g), the shifts in the peak energy cannot be an artefact due to the presence of holes, as the shifts are too large relative to the ground-state linewidth. Moreover, in general the process of photoexcitation instantaneously creates holes symmetrically within the in-homogeneously broadened gaussian distribution of electrons in the VB for any given $k$-vector. This is confirmed in our experiments at high photoexcitation intensity (Fig. 2g), as well as at low photoexcitation intensity (see SI figure 5). Thus, the shift in peak energy – measured at the instant of photoexcitation, cannot be an artefact due to holes, as that would require an asymmetric hole distribution.

**Quantitative Comparison to Microscopic Theory**

To obtain a microscopic understanding and a quantitative comparison with experiment, we perform *ab initio* time-dependent adiabatic GW (TD-aGW) calculations[35,36] of 1L WS$_2$ (Fig. 3a-c). In this approach, the single-particle density matrix $\rho_{mnk}(t)$, where $m$ and $n$ are band indices and $\mathbf{k}$ a wavevector, is propagated in time under the presence of an external field E(t), which simulates a short, 50 fs optical pump that resonantly creates an initial exciton population. We then propagate the electronic Hamiltonian and compute the electronic spectral function in the presence of the excitonic population. In the linear-response limit, this method yields optical absorption spectra equivalent to state-of-the-art formalisms based on interacting Green's function formalism, notably the first-principles GW plus Bethe-Salpeter equation (GW-BSE) approach[37]. Importantly, beyond the linear-response regime, this real-time formalism allows us to include many-electron interactions.

In Fig. 3d, we see very good agreement between our experiment and theory comparing the shift of the valence band maxima from the center of the K-valley ($\Delta k_{max}$) (see supporting material) with increasing exciton density. As discussed above, the appearance of the VBM at a finite wavevector represents the appearance of the Bogoliubov dispersion[16], and is a key signature of the BEC-BCS transition for the excitonic state. We also compare the decrease in energy at the K point ($\Delta E_X$) with increasing exciton density (Fig. 3e), observing a linear dependence with a shift of around 80 meV shift for the highest density measured. Overall, the results of our *ab initio* TD-aGW calculations (Fig. 3a-c) agree well with the experimental TR-ARPES (Fig. 2a-d) measurements across an order of magnitude of exciton density and capture the notable features in experiment discussed above.

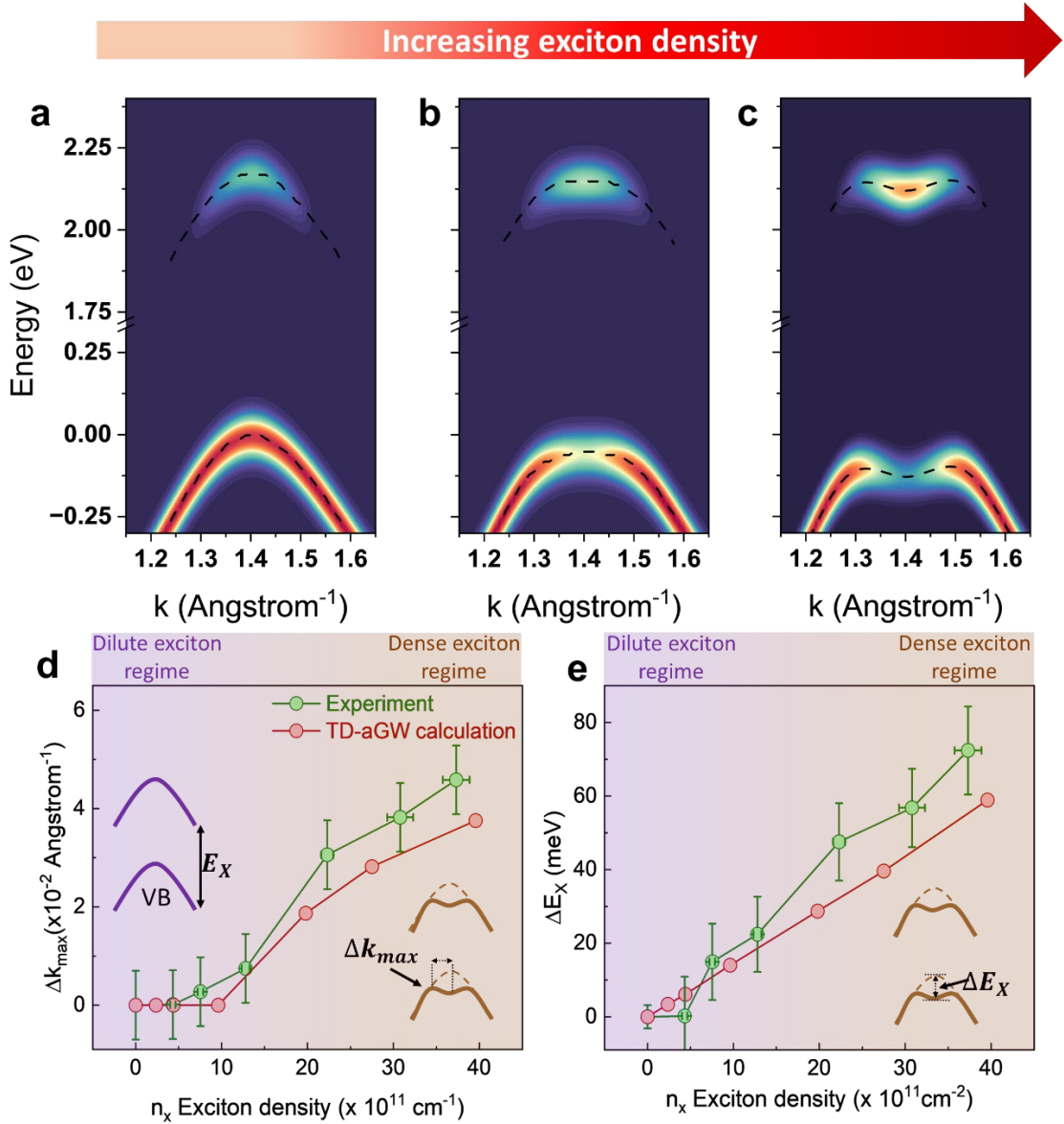

**Figure 3. TD-aGW calculation.** (**a-c**) Energy-momentum ARPES spectra of the valence band and the exciton bound electron in 1L WS$_2$ with increasing exciton density as calculated using ab initio time-dependent adiabatic GW methods. The calculations we performed for a range of mid-$10^{11}$ to high-$10^{12}$ cm$^{-2}$ exciton densities. (**d**) The shift in the valence band maximum (VBM) as a function of exciton density. In the low density, the VBM remains at the high symmetry point at K. For high densities, the VBM shift away from the center of the valley by a finite amount denoted by $\Delta k_{max}$. The experiment and theory also reproduce the classic spectroscopic features associated with the BEC (dilute regime) and BCS (dense regime) phases in a non-equilibrium excitonic insulator. (**e**) The magnitude of the downward shift in the valence band energy peak at the K point ($\Delta E_X$) increases linearly with increasing exciton density.

**Connection with exciton-driven Floquet effect**

Finally, we relate our experimental findings to the recently predicted giant exciton-driven Floquet effect [17].

In recent years, implementations of the Bloch-Floquet formalism have shown that one can create nontrivial phases of matter by driving systems with a time-periodic perturbation[38,39]. To date, most such studies have utilized an external light field oscillating at a frequency $\omega_p$ to provide the time-periodic interactions[40–46]. Therein, one obtains a time-dependent coupling between an occupied state in the valence band and an unoccupied state in the conduction band, given by the interaction Hamiltonian $\sim e\mathcal{E} \cdot d\; e^{-i\omega_p t}$ , where $\mathcal{E}$ and $d$ are the magnitude of the electric field of light and the dipole matrix element between the valence and conduction band states, respectively (Fig. 4a). Nonetheless, the use of optical fields to implement Floquet physics presents some formidable challenges, such as unwanted heating, absorption, and sample damage, due to the high-intensity requirements of the optical pulse[21].

Another powerful idea that has recently emerged is to use excitons to provide the time-periodic perturbation[17] . In the dense exciton limit, the change in the quasiparticle self-energy, $(\delta\Sigma^X)$, when analyzed in the time-domain, provides exactly such a time-dependent interaction. Oscillating with the phase $e^{-i\omega_X t}$, where $\omega_X = E_x/\hbar$ is the exciton frequency, $\delta\Sigma^X$ provides a coupling between conduction and valence band states and gives rise to the same effective Floquet physics obtained via a light-driven periodic perturbation (Fig. 4b). In 2D semiconductors, $\delta\Sigma^X$ , being proportional to the square root of the exciton density $\sqrt{n_X}$ and the exciton binding energy[17], offers a significantly stronger time dependent coupling in the dense exciton limit: the weak dielectric screening results in large exciton binding energies and also allows for higher exciton densities prior to the Mott limit. A strong time dependent Floquet perturbation creates new opportunities for engineering Floquet states. In particular, it allows for the hybridization between dressed and bare states that is critical to the implementation of Floquet engineering, but has only rarely been achieved [40,45]. In our experiments, the observed Mexican-hat-like dispersion of the valence band at the highest exciton densities is interpretable exactly as the Floquet state formed due to the hybridization between the bare valence and exciton-dressed conduction band (Fig. 4b).

To make a quantitative comparison, we consider the Floquet effects due to an exciton population generated by an ultrafast optical pulse and compare them to the optically driven Floquet effects by an identical ultrafast pulse. In particular, we compare the Floquet hybridization magnitude represented by $\Delta$E (see Fig. 4a, b) that can be achieved under the two schemes given the same input resources. From our microscopic analysis and previous literature[17], the change in the electron-hole self-energy coupling matrix elements in Fourier space can be expressed as $\delta\Sigma^X_{vck} = E_b A_{vck}\sqrt{n_X}$, where $E_b$ is the exciton binding energy, and $v$ and $c$ label valence and conduction band states, respectively, at wavevector $\boldsymbol{k}$ (see Methods). For a 2D Wannier exciton localized in k-space around the VBM, the overall magnitude of the hybridization is then proportional to $\Delta E_X \equiv$ $|\delta \Sigma^X_{vck}|^2 / E_b$. Thus, for an exciton density of $n_X \sim 5 \times 10^{12}$ cm$^{-2}$, we obtained a hybridization magnitude $\Delta E_X \sim 100$ meV, consistent with our experimental measurements (Fig. 3e).

On the other hand, for optically driven Floquet, the coupling matrix elements can be expressed as $\delta \Sigma^{opt}_{vck} \sim e \, \mathcal{E} \cdot d_{vck}$, where $e$ is the electron charge, $\mathcal{E}$ is the amplitude of the external electric field potential, and $d_{vck}$ are interband optical transition matrix elements. To maximize the impact of optical-driven Floquet in a practical experiment, we seek to tune the optical pulse as close to the band edge as possible, while also ensuring that we avoid mid-gap absorptive regions. In existing TMDC samples, there is substantial absorption above the A exciton resonance, due to higher order excitonic states and other excitonic species[47,48]. Hence the highest energy that one can tune to while avoiding absorption is just below the A-exciton peak. Under these conditions, the magnitude of the hybridization is $\Delta E_{opt} \equiv |\delta \Sigma^{opt}_{vck}|^2 / E_b$, For a 50 fs pulse with electric field amplitude of $\sim 4 \times 10^4$ $V cm^{-1}$ we estimate a hybridization magnitude $\Delta E_{opt} \sim 1$ meV for optical Floquet.

We confirm this analysis experimentally by pumping our sample with $5 \times 10^4$ $V cm^{-1}$ pulse at 1.98 eV – 120 meV below the A exciton resonance – to avoid absorption in the monolayer. We chose s-polarized light to avoid Volkov states[42,49] (see supporting material). Figure 4c shows the measured ARPES spectrum at zero ps pump-probe delay, displaying a clear, though weak, replica of the valence band 1.98 eV higher, but no observable hybridization effects. This is consistent with our theoretical expectation ($\Delta E_{opt} \sim 1$ meV), and experimental resolution.

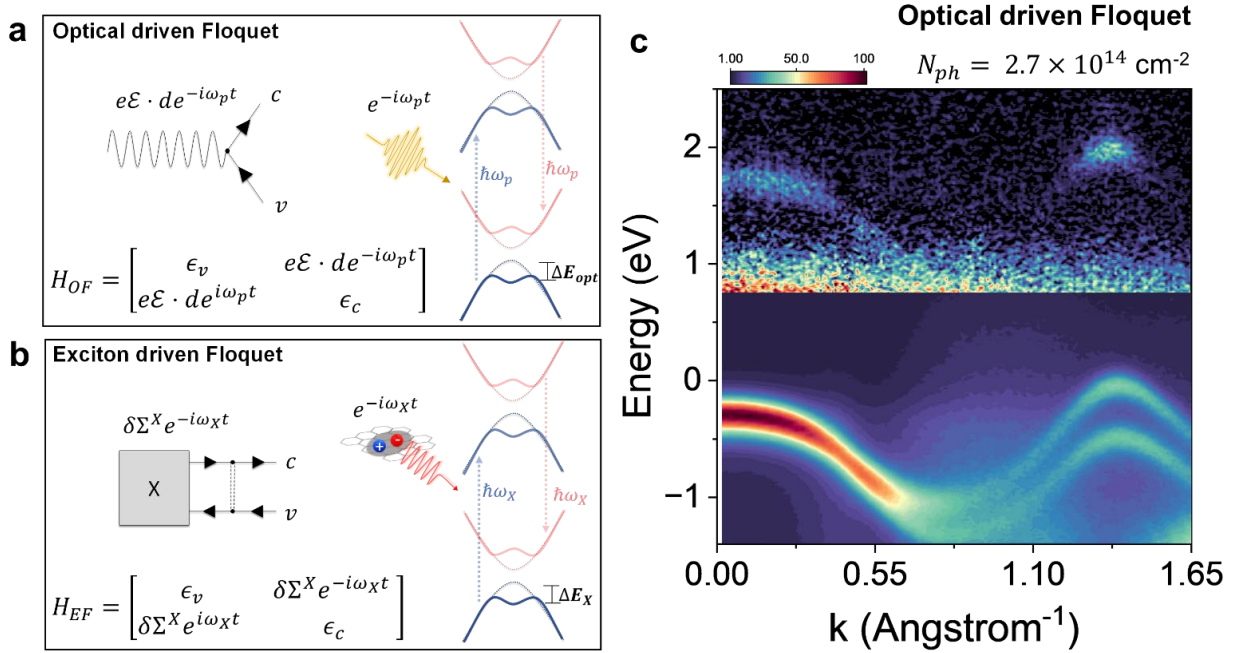

**Figure 4. Optical and Exciton driven Floquet effects.** Feynman diagrams, effective Hamiltonian, and schematics of the (hybridized) replica bands for (**a**) the optically driven and (**b**) the exciton driven Floquet processes. $e \, \mathcal{E} \cdot d \, e^{-i\omega_p t}$ and $\delta \Sigma^X e^{-i\omega_X t}$ denote the time-periodic perturbations in the respective cases; $c$ and $v$ denote electrons in the conduction, valence band respectively, with $\epsilon_c$ and $\epsilon_v$ their corresponding energies; And $\Delta E_{opt}$ and $\Delta E_X$ denote the magnitude of hybridization in the two cases. (**c**) ARPES spectrum for 1L WS$_2$ driven by an intense optical pulse 120 meV below the exciton resonance. For a peak intensity 2.5 $\times 10^9$ $W cm^{-2}$ (peak field strength: $5 \times 10^4$ $V cm^{-1}$), we observe weak Floquet replicas over the entire Brillouin zone, but no band hybridization effects.

## Conclusion

The direct experimental observation of the non-trivial electronic structure induced by an intense excitonic field opens up many new avenues of investigation. To begin with, this suggests that intense bosonic fields (in addition to photons) – such as excitons, phonons, magnons and plasmons, could indeed be utilized to drive new quantum phases and topologies in materials[16,17,20,50,51], with each bosonic mode offering its own unique advantages and opportunities. The observation of the features of the BCS state in a photoexcited 1L TMDC triggers analogies with superconductivity and superfluidity, and potential applications therein[1,15,52]. Similarly, resonantly and coherently photoexcited TMDCs now offer a new and powerful platform for the investigation of XI and NEQ-XI. The availability of the bandstructure in the absence of excitons, the ability to control exciton density and the absence of other coincidental phase transitions could help unambiguously access the properties of the macroscopic excitonic population and its influence on the electronic structure of a material. Arguably, studies on conventional XI to date have suffered some ambiguity and controversy due to the unavailability of these features[53,54].

Our work also provides a powerful new approach towards Floquet engineering using excitonic fields instead of optical fields. Excitonic fields also avoid the detrimental and competing effects associated with optically driven Floquet, including unwanted absorption and laser-assisted photoemission effect (LAPE)[21,49]. Finally, the significantly stronger Floquet effects using excitonic fields, could enable Floquet engineering proposals in 2D materials and semiconductors that have remained out of reach using optical drivers.

# Methods

## Sample preparation

The WS$_2$ monolayer was grown at IMRE (Singapore) on a sapphire substrate using chemical vapor deposition in a PlanarTECH quartz tube furnace at a temperature of 900°C and atmospheric pressure in a forming gas environment[55]. Large area sheets (~mm size) of monolayer WS$_2$ were enabled by a patented custom-designed crucible[56]. Micron scale single domain regions were chosen for measurement as described below. Bulk hexagonal boron nitride flakes (~25 nm thick) were cleaved on low resistivity n-doped silicon wafers using scotch tape. The WS$_2$ sheet was then transferred to the cleaved hBN using a water-assisted transfer method with PMMA/PDMS polymer stack. After a successful transfer, acetone and isopropanol alcohol were used to clean the transferred polymer. The WS$_2$/hBN stack was then heated at ~300°C for ~10 hours in UHV before the ARPES experiments.

## Experimental details

Time-resolved XUV micro-ARPES experiment was performed in a time-of-flight momentum microscope coupled with a home-built table-top higher harmonic generation setup. The table-top high harmonic setup was driven by a Yb-doped fiber laser system operating at 2 MHz with 230fs pulse width. The laser system delivered $50\mu J$ pulse energy out of which 35 $\mu J$ (1030 nm) was frequency doubled by a thin BBO to generate ~15 $\mu J$ of 515 nm. The 515nm laser was then focused onto a Kr gas jet to generate the XUV probe (21.7 eV). The XUV probe was finally collected and focused onto the sample using an ellipsoidal mirror, providing a spot size ~60 $\mu m \times 6\ \mu m$ at the sample. The remaining power from the laser ($15\mu J$ of 1030 nm) was used to pump a non-colinear optical parametric amplifier (NOPA) to generate tunable pump excitation. For the A exciton resonant excitation, a 2.1 eV s-polarized pump was used with varying intensity. For the optical Floquet experiment, a below-band gap excitation of 1.98 eV was used for both the s-and p-polarizations. The XUV probe was p-polarized for the entire set of experiments.

The time-of-flight momentum microscope uses an immersion objective lens to collect the photoemitted electrons from the sample. The microscope has <30 meV energy resolution and <0.01 Å$^{-1}$ momentum resolution. An ~16 × 16 $\mu m^2$ area of a clean single domain monolayer region of the sample was selected by inserting a field aperture (16 $\mu m$) in the image plane of the lens to perform spatially resolved ARPES measurement. The instrument was then switched to the momentum mode to allow the photoemitted electrons within the aperture to pass through the drift tube and finally get collected by a time-of-flight detector resolving their energy and momentum. All the ARPES experiments were performed at 100K.

## *Ab initio* GW and GW-BSE calculations

Density-functional theory (DFT) calculations are performed with the Quantum Espresso package[57] within the local density approximation (LDA) to the exchange-correlation functional. We use ONCVPSP norm-conserving LDA pseudopotentials[58] with the W 5s, and 5p semi-core states included as valence states. For the ground-state calculation and the Kohn-Sham orbitals of monolayer $WS_2$, we use a **k**-point grid of 12×12×1 and a plane wave energy cutoff of 50 Ry. A vacuum of 15 Å in a supercell and truncated Coulomb interaction are employed to prevent spurious interactions between periodic images. The GW and GW plus Bethe-Salpeter equation (GW-BSE) calculations are carried out with the BerkeleyGW package[59–61]. A **k**-point grid of 6×6×1 with a subsampling of 10 points in the volume of the Voronoi cell around the **q**=0 point[62], and a dielectric energy cutoff of 25 Ry and 1800 bands are used in GW calculation. The frequency dependence of the dielectric screening is computed using the generalized plasmon pole model[58]. The direct Kohn-Sham band gap in our DFT calculation is 1.60 eV, and the GW quasiparticle band gap is 2.62 eV. The BSE kernel is computed on a **k**-point grid of 36×36×1 with 6 lowest conduction bands and 6 highest valence bands.

## *Ab inito* time-dependent adiabatic GW (TD-aGW) calculations

We detail below the procedure to obtain the spectral properties of the system within the *ab inito* time-dependent adiabatic GW (TD-aGW) calculations in monolayer $WS_2$. In this approach, the single-particle density matrix $\rho_{vck}(t)$, where $v$ denotes a valence band, $c$ a conduction band, and **k** a wavevector, is propagated in time under the presence of an external field E(t), which simulates a short, 50 fs optical pump, that resonantly creates an initial exciton population. After the external field vanishes, the evolution of the quasiparticle in the system, including many-electron interactions and nonequilibrium occupations, can be rigorously expressed through a time-dependent Hamiltonian

$$H(t) = H^{QP} + \delta\Sigma(t),$$

where $H^{QP}$ is the equilibrium quasiparticle Hamiltonian, which describes electronic excitations including many-body renormalizations in the electronic ground state, and $\delta\Sigma(t)$ is the sum of the change in the many-electron screened-exchange interaction and the change in the Hartree term. For weak external illumination, $\delta\Sigma(t)$ depends primarily on the many-electron screened-exchange interaction, which depends on the photoinduced change of the occupation of electrons at the VBM and CBM (since the exchange only involves interactions with occupied electronic states). Higher-order effects, such as the photoinduced change in screening, play a smaller role, and may be neglected in our analysis.

The equation of motion for the density matrix, $-i\hbar\partial_t\rho_{nmk}(t) = [H^{eff}(t), \rho(t)]_{nmk}$, is given effectively by the Hamiltonian

$$H^{eff} = \sum_{\mathbf{k}} \begin{pmatrix} \epsilon_{v\mathbf{k}} + \delta\Sigma_{vv\mathbf{k}}(t) & \delta\Sigma_{vc\mathbf{k}}(t) \\ \delta\Sigma_{vc\mathbf{k}}^*(t) & \epsilon_{c\mathbf{k}} + \delta\Sigma_{cc\mathbf{k}}(t) \end{pmatrix},$$

where $\epsilon_{v\mathbf{k}}$ and $\epsilon_{c\mathbf{k}}$ are the equilibrium quasiparticle energy of states in the valence (hole) and conduction (electron) band with wavevector $\mathbf{k}$, respectively (only two bands are written for simplicity), $\delta\Sigma_{mn\mathbf{k}}(t) = \sum_{ij\mathbf{k}'} W_{mn\mathbf{k},ij\mathbf{k}'}\,\rho_{ij\mathbf{k}'}(t)$ is the effective and time-dependent electron-hole interaction involving bands $m$, $n$, $i$ and $j$, and $W$ are matrix elements of the screened Coulomb interaction. The dielectric function is fixed during the calculation. The optical response of such formalism, within the linear response limit, is equivalent to that obtained within the well-established ab initio GW-BSE approach, as proved, for instance, in Refs.[35,63,64]. Notably, when $\rho_{vc\mathbf{k}}(t)$ correspond to the self-consistent change in the density matrix caused by an exciton, we can express the change in the self-energy in Fourier space, $\delta\Sigma_{mn\mathbf{k}} = E_b A_{mn\mathbf{k}}\sqrt{n_X}$, where $E_b$ is the exciton binding energy, and $A_{mn\mathbf{k}}$ is the exciton envelope function.

The results of our *ab initio* TD-aGW calculations (Fig. 3a-c) agree well with the experimental TR-ARPES measurements across an order of magnitude of exciton density and capture the notable features in experiment discussed above. Moreover, our microscopic analysis immediately gives us qualitative insights into the physical origin of such effects. For the range of exciton densities we study, the off-diagonal coupling $\delta\Sigma_{vc\mathbf{k}}(t)$ – which is proportional to $\sqrt{n_X}$ (see following section) – is larger and dominates over the diagonal coupling terms $\delta\Sigma_{vv\mathbf{k}}(t)$ and $\delta\Sigma_{cc\mathbf{k}}(t)$ – both of which are proportional to the exciton density $n_X$. Hence, the band renormalization effects we observe are not explained simply as photodoping but are in fact much better described in terms of a time-dependent coupling which, just as in the case of optically driven materials, can be interpreted in light of a Floquet formalism.

The connection between the exciton-driven Floquet formalism and the effective Hamiltonian $\boldsymbol{H^{eff}(t)}$ is most easily seen in the linear regime: in frequency space, the linear response theory of $\boldsymbol{H^{eff}(t)}$ becomes an eigenvalue problem known as the Bethe-Salpeter equation (BSE) that yields excitons with energies $\hbar\omega_X$. Hence, in the time domain, excitons correspond to oscillations of the density matrix with a time dependence $\boldsymbol{\rho_{vc\mathbf{k}}(t) \sim e^{-i\omega_X t}}$. This allows for the abovementioned Floquet interpretation, where the electron-hole coupling acquires a time dependence $\boldsymbol{\delta\Sigma_{vc\mathbf{k}}(t) \sim e^{-i\omega_X t}}$. Hence, it is no surprise that, in the low-density limit (Figs. 2a and 3a), the replica states obtained from such a Floquet picture are equivalent to the excitonic states obtained from the solution of the BSE. This explains why the photoemission signature from the Floquet replica band in the low density is equivalent to the earlier reported photoemission signature from electrons bound to excitons in TR-ARPES experiments.

## Time-dependent angle-resolved photoemission intensity

To compute the time dependence of the angle-resolved photoemission intensity after the pump field, the central quantity is the lesser component of two-time Green's function $G^<(t, t')$. With an instantaneous self-energy (as the case with the TD-aGW approximation), $G^<(t, t')$ can be computed from the retarded component of the Green's function $G^R(t, t')$, the advanced component of the Green's function $G^A(t, t')$, and the density matrix $\rho(t)$ using[65],

$$G_k^<(t, t') = -G_k^R(t, t')\rho_k(t') + \rho_k(t)G_k^A(t, t')$$

The inverse ARPES which measures the unoccupied (quasielectrons) is computed from $G^>(t, t')$ with a similar expression

$$G_k^>(t, t') = -G_k^R(t, t')\overline{\rho_k}(t') + \overline{\rho_k}(t)G_k^A(t, t')$$

Where $\overline{\rho_k} = 1 - \rho(t)$. For instantaneous self energy, the retarded component is written as

$$G_k^R(t, t') = -i\theta(t - t')\mathcal{T}e^{-i\int_{-t}^{t} H(\overline{t})d\overline{t}}$$

Where $\mathcal{T}$ is the time ordering operator, while the advanced component is obtained by $G_k^A(t, t') = G_k^R(t', t)^\dagger$

## Methods References

# Acknowledgements

The authors thank the Okinawa Institute of Science and Technology Graduate University (OIST) engineering support section and thank Y. Yamauchi from the OIST Facilities Operations and Use section for their support. V.P. and K.M.D. thank O. Karni for discussions. The authors also thank

M.G. Menezes for discussions. V.P., D.R.B., X.Z., N.S.C., J.P.U., M.K.L.M., J.M., K.M.D. acknowledge that this work was supported by the Femtosecond Spectroscopy Unit of OIST. M.K.L.M. acknowledges the support from JSPS Kakenhi grant no. 17K04995. J.M. acknowledges the support from JSPS Kakenhi grant no. 21H01020. K.M.D. acknowledges the support from JSPS Challenging research Pioneering grant no. 22K18270. The authors acknowledge the XUV generation technology was supported by the OIST Innovative Technology – Proof of Concept (POC) Program. K.M.D. and N.S.C. acknowledge the support from Kick-start fund KICKS-OIST. F.H.J. acknowledges the support by the Center for Non-Perturbative Studies of Functional Materials Under Non-Equilibrium Conditions (NPNEQ), funded by the US Department of Energy (DOE) Office of Science under contract DE-AC52-07NA27344. This work uses codes developed by the Center for Computational Study of Excited State Phenomena in Energy Materials (C2SEPEM), which is funded by the DOE Office of Science under Contract No. DE-AC02-05CH11231. The authors acknowledge the support for computational resources and storage provided by National Energy Research Scientific Computing Center (NERSC), a DOE Office of Science User Facility under Contract No. DE-AC02-05CH11231, and the National Center for High-performance Computing (NCHC). D.Y.Q. acknowledges support by the U.S. Department of Energy, Office of Science, Basic Energy Sciences, under Early Career Award No. DE-SC0021965. F.B. and K.E.J.G. acknowledge funding support from Agency for Science, Technology and Research (#21709) and a Singapore National Research Foundation grant (CRP21-2018-0001). K.W. and T.T. acknowledge support from JSPS KAKENHI (Grant Numbers 19H05790, 20H00354 and 21H05233).

## Author contributions

V.P., D.R.B. and X.Z. performed the TR-ARPES measurements, the sample characterization and analysis of the data, with assistance from J.P.U., M.K.L.M. & J.M., under the supervision of K.M.D. Y.H.C. implemented the theoretical simulations, with assistance from D.Y.Q., and supervision of F.H.J. K.E.J.G. developed the CVD large monolayer $WS_2$ growth process. F.B. prepared the samples with assistance from N.S.C. & V.P., under the supervision of K.E.J.G. and K.M.D. K.W. and T.T. provided bulk hBN crystals for the study. F.H.J. and K.M.D. conceived and designed the study. All authors contributed to the writing and reviewing of the manuscript.

# Supplementary Information

## Absorption spectroscopy and Photoemission excitation characterization

The monolayer $WS_2$ sample was characterized using absorption spectroscopy before the ARPES experiment at sub 100K. The sample showed strong absorption at ~2.1 eV (590 nm), as seen in Fig 1d. Before the TR-ARPES experiments in the dense exciton limit, the pump excitation wavelength was tuned around the absorption peak to confirm and identify the exciton resonance carefully. A constant pump power of ~10 mW was maintained for all pump wavelengths and the photoemission signal was normalized with respect to the total photoemission counts to remove any variations between the experiments. The photoemission signal from the exciton-bound electrons showed maximum intensity for ~2.1 eV pump excitation, which is resonant to A-exciton in 1L $WS_2$ as seen in Fig. 1e.

## Dispersion fitting

### Valence band dispersion fitting

We obtain a three-dimensional data set resolved in E, $k_x$, and $k_y$ from the ToF-MM, which is then corrected for distortion using the procedure outlined in reference[25]. We use the following procedure to track changes to the valence band dispersion due to the presence of excitons – The distortion-corrected 3D dataset is dissected along the $\Gamma - K - M$ high symmetry points of Brillouin zone by integrating 7 k-pixels in the perpendicular direction (Fig. 1c inset). A small 2-dimensional E-k region ($k_{ll}: \pm 0.2\ \text{Å}^{-1}; E: +0.5\ to - 1.2\ eV$ ) is isolated around the top of the valence band at K from the rest of the band structure. For each $k_{ll}$ position, we obtain a distribution in energy which is generally referred to as energy distribution curve (EDC). SI figure 1a shows the EDC from the valence band in the -1.2eV to +0.5 eV range (the top of VB1 is set as 0 eV), with the two peaks corresponding to VB1 and VB2 at the K point. A Shirley-type background is subtracted from the two peaks to improve the peak fitting accuracy. A double gaussian function is used to fit the two peaks together (SI figure 1b) for the range of $k_{ll}: \pm 0.2\ \text{Å}^{-1}$ around the K point.

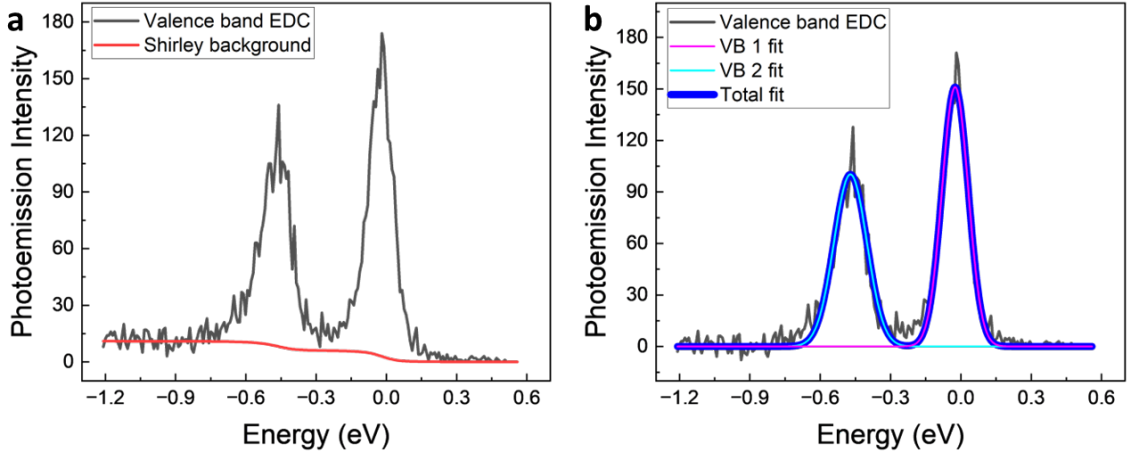

***SI figure 1. Background correction, and double Gaussian fitting: (a)*** *The valence band EDC at the K point with the calculated Shirley background,* ***(b)*** *Valence band EDC after subtraction is fitted to a double gaussian function to get the peak position for both VB1 and VB2 to calculate the exciton-driven hybridization effect.*

This process is repeated for all the different photoexcitation densities shown in Fig. 2. For the A-exciton resonant excitation, the band dispersion at the edge ($k_{ll} < -0.15$ Å$^{-1}$ with respect to the K point) remains unchanged which allows us to correct for the shift of VB1 and VB2 caused by the surface photovoltage effect for all the different exciton densities. The dispersion of VB1 and VB2 for all the different photoexcitation condition are shifted appropriately to match the dispersion of the un photo-excited dispersion of VB1 and VB2 at the edge ($k_{ll} < -0.15$ Å$^{-1}$ with respect to the K point) to compare the change in the dispersion and calculate $\Delta E_X$. SI figure 2(e-h) compares the band dispersion of VB1 and VB2 for the un-photoexcited and the photoexcited cases after correction. Notice photoexcited VB2 dispersion shows no change compared to the unperturbed VB2 dispersion.

**Exciton bound electron dispersion fitting**

We process the exciton bound electron photoemission signal at ~2.1 eV for all the exciton densities similarly by dissecting the 3D data set along the $\Gamma - K - M$ high symmetry axis by integrating 7 k pixels in the perpendicular direction. For each $k_{ll}$ position around K point ($\pm 0.2$ Å$^{-1}$) we fit the EDC to a single gaussian function to obtain the dispersion as a function of k around the K point. SI figure 2(a-d) shows the obtained dispersion of the replica state as a function of decreasing exciton density.

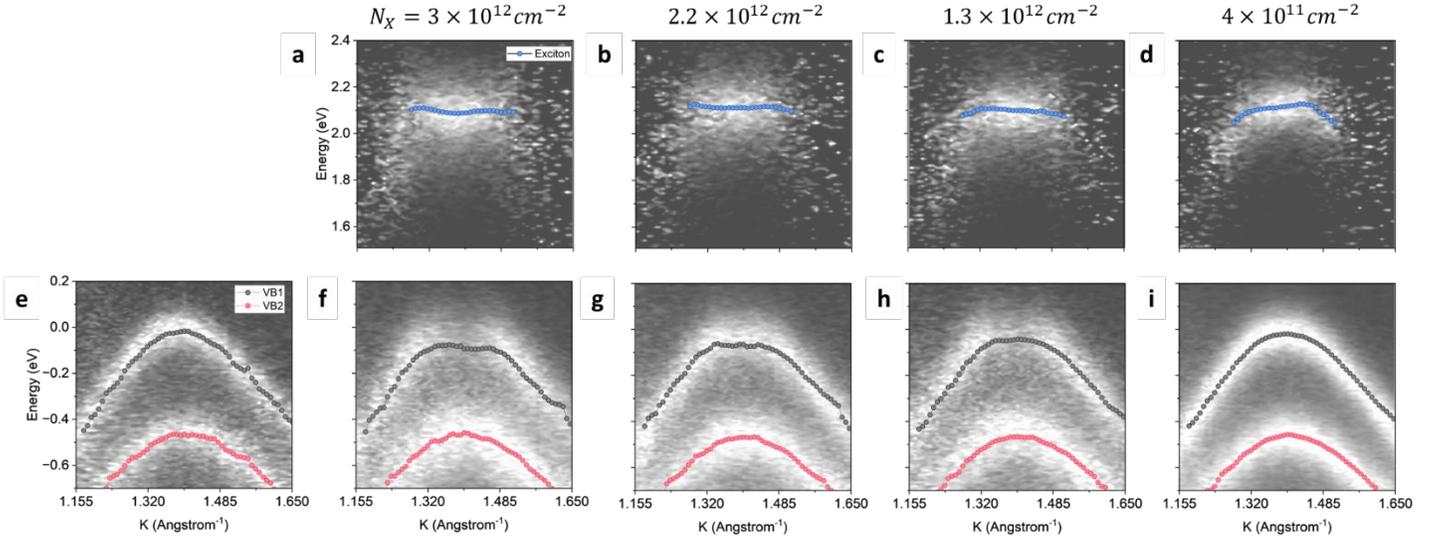

**SI figure 2. Fitted dispersion for the valence bands and the exciton bound electron: (a-d)** *Dispersion of the valence band replica with decreasing exciton densities obtained by single gaussian fitting.* **e)** *Dispersion of VB1 and VB2 for no pump case for comparison* **(f-i)** *Dispersion of VB1 and VB2 with decreasing exciton densities obtained by fitting a double gaussian function to the Shirley-corrected EDCs.*

## Calculation of magnitude of downward shift of VB at K ($\Delta E_X$) and valence band maxima ($\Delta k_{max}$)

We first follow the procedure mentioned in the valence band fitting section to remove any shift in photoexcited VB1 and VB2 dispersions, caused by the surface photovoltage effect. We then match the VB1 and VB2 dispersion for the non-photoexcited and the photoexcited case away from the K point ($k_{ll} < -0.15 \text{ Å}^{-1}$). We then subtract the photoexcited VB1 dispersion from the non-photoexcited dispersion VB1 dispersion to get the $\Delta E_X$ curve for different exciton densities as shown in SI figure 3a. We repeat the same for VB2 as shown SI figure 3b. We clearly observe a significant change in the VB1 dispersion with increasing exciton densities and no change is observed in the dispersion of the VB2. To calculate the magnitude of $\Delta E_X$, we take the average of the change in VB1 around a small section near the K point ($\pm 0.02 \text{Å}^{-1}$) for each exciton density and plot it against the exciton density in Fig. 3e.

To obtain the valence band maxima, we consider the VB dispersions obtained earlier as a function of exciton density. We choose the $k_{ll}$ value from the unperturbed valence band with the maximum energy as the reference and subtract it from the $k_{ll}$ positions obtained for the maximum energy from the dispersion for different exciton densities. This relative shift of the valence maxima is plotted as a function of exciton density in Fig. 3d which matches well with the changes obtained for the theoretical calculations.

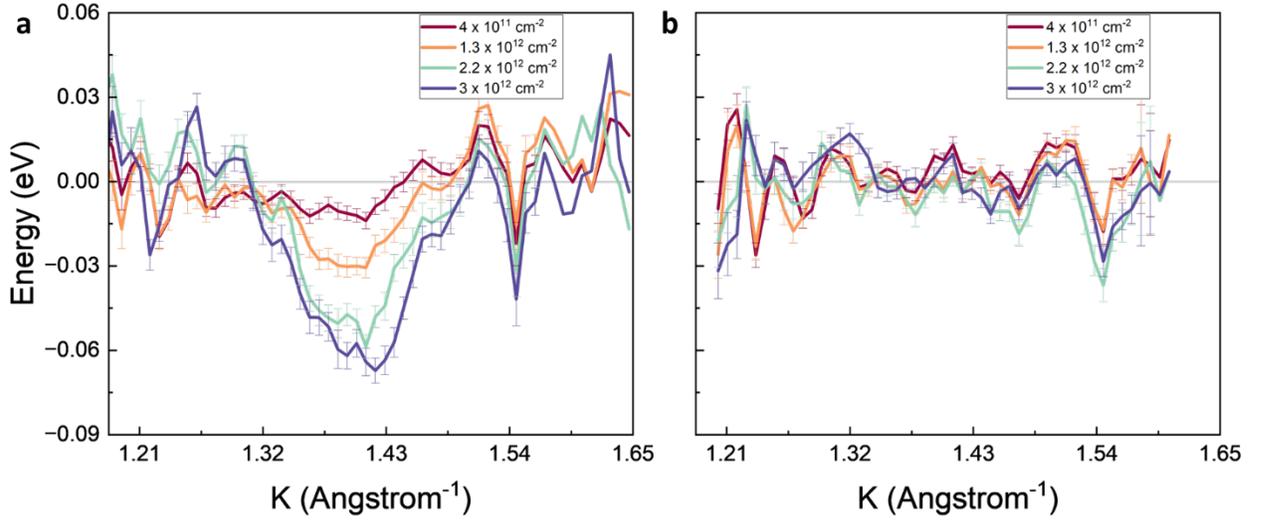

***SI figure 3. Magnitude of downward shift of the valence band as a function of exciton density: (a)***
*Difference in the dispersion for VB1 for different exciton densities with respect to the no pump VB1 dispersion
showing drastic changes near the K point, (**b**) Difference in the dispersion for VB2 with respect to the no
pump VB2 dispersion showing no significant change in general.*

## Hole density calculation

To calculate the hole density from the TR-ARPES data, we focus on the top of the valence band,
i.e., VB1 only. We compare the photoemission intensity between the un-photoexcited and
photoexcited band structure for the VB1, as shown in SI figure 4. First, we choose a small k space
region around the top of the VB ($0.15 \text{ Å}^{-1} \times 0.15 \text{ Å}^{-1}$) and integrate the corresponding energy
distribution curves (EDC) within a grid of $3 \times 3$ $k$-pixels (or $0.02 \text{ Å}^{-1} \times 0.02 \text{ Å}^{-1}$) to improve the
signal-to-noise ratio. We then choose a small energy range at the top of the valence band and
integrate the momentum image within this range to obtain the energy-integrated momentum
images of the top of VB1. This process is done for both the un-photoexcited and the photoexcited
VB1. The next step is to correct, if any, momentum offsets between the two measurements. This
is done by fitting the EDC at each $[k_x, k_y]$ pixel. The energy range for the EDCs is truncated to
only include VB1 and VB2 to prepare them for fitting, following similar procedure as mentioned
in the valence band fitting section. Next, we use the Shirley background removal technique to
remove the background in the photoemission spectra. This allows us to track the changes in the
VB1 and VB2 peaks more accurately (see SI figure 1b). Next, the two peaks are fitted using a
double gaussian function to extract the peak position (dispersion), the width, and the area under
the curve. The VB1 dispersion (peak position) is then fitted with a 2-D paraboloid to obtain the
peak position of the top of the valence band for the two cases, and any offset obtained in this
process is eliminated by shifting one of the valence bands as required. The next step is eliminating
the intensity variation between the measurements due to experimental conditions. For this, the data is normalized by the intensity in a small k-space region around $[k_{x0}, k_{y0}]$ (marked in the red box in SI figure 4) for both measurements. The region is unaffected by the pump excitation as it is far from the hole region. This now allows us to calculate the hole occupancy distribution, which is given by

$$f[k_x, k_y] = 1 - \frac{I_{\text{unphotoexcited}[k_{x0}, k_{y0}]} I_{\text{photoexcited}[k_x, k_y]}}{I_{\text{photoexcited}[k_{x0}, k_{y0}]} I_{\text{unphotoexcited}[k_x, k_y]}}$$

Finally, the hole density is given by multiplying the hole occupancy distribution with the density of hole states in momentum space and summing over the entire distribution

$$n_{hole} = \frac{2}{L^2} \iint f[k_x, k_y] \frac{dk_x dk_y}{\frac{4\pi^2}{L^2}}$$

$dk_{x,y} = 0.0066$ Å$^{-1}$ is the k-space coordinate spacing in the $k_{x,y}$ directions. The pre-factor 2 is for considering the two K valleys in the BZ.

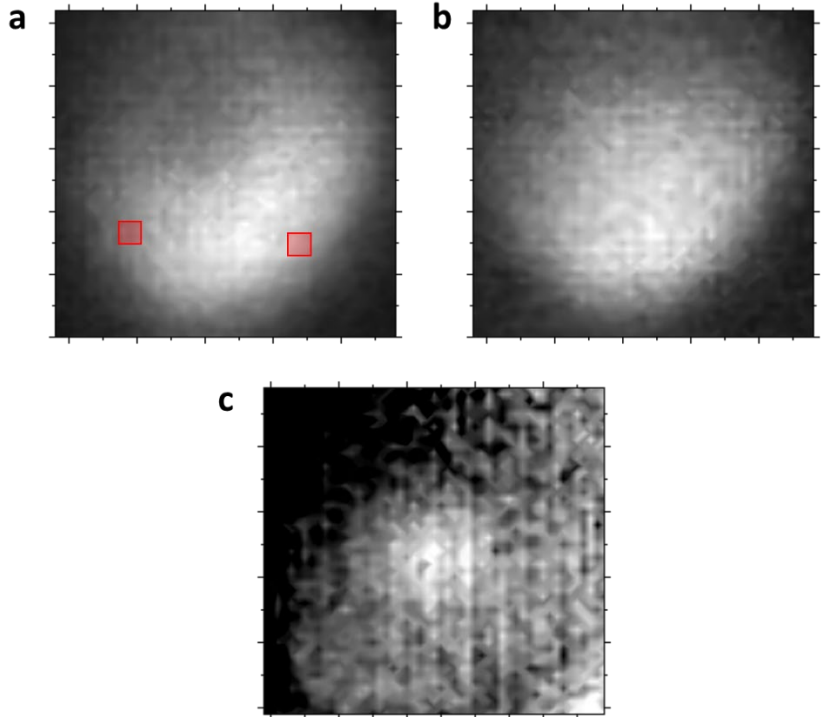

*SI figure 4. Exciton density calculation:* *(a)* *Energy integrated momentum image of the top of the K valley (VB1) with an exciton density* $n_X = 3 \times 10^{12} \, cm^{-2}$. *The red boxes indicate the region used to normalize the intensity for the un-photoexcited and photoexcited case.* *(b)* *Energy integrated momentum image of the top of the K valley for the un-photoexcited case.* *(c)* *The momentum distribution of the hole wavefunction obtained after taking the ratio of photoexcited and un-photoexcited bands in* *(a)* *and* *(b)* *respectively.*

## Low exciton density energy distribution curves

We confirm the uniform photoexcitation of exciton bound holes by observing the EDCs around the K valley for the low exciton density ($n_X = 4 \times 10^{11} \mathrm{cm}^{-2}$). In this limit, we do not observe any shift in the spectra and hence can directly compare the loss in the intensity due to the photoexcited exciton bound holes with the non-photoexcited case. As seen in SI figure 5, the EDC at Kx = 0 shows reduced uniformly across the entire spectrum.

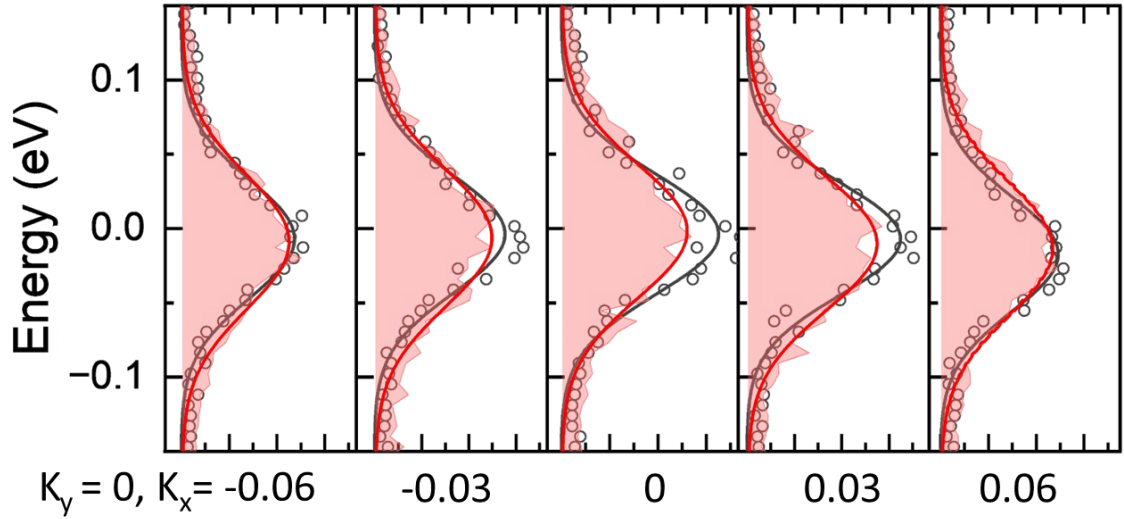

**SI figure 5. Uniform excitation of exciton bound holes:** *Energy distribution curves of the valence band maxima at K valley for low exciton density (red filled curves) and the non-photoexcited valence band maxima at K valley (black circles). As seen clearly at the K point the loss in the signal in the low exciton density case is uniformly distributed in energy. The solid lines are gaussian fits to the data.*

## Background correction for ARPES data

To clearly visualize the optical Floquet replica, the ARPES data was corrected to remove the background signal from the extended tail of the valence band. The background signal was removed using an asymmetric least square smoothing algorithm. SI figure 6 shows the ARPES data before and after the correction for the optical Floquet replicas, where the signal from the replica are unaffected but can be seen more clearly.

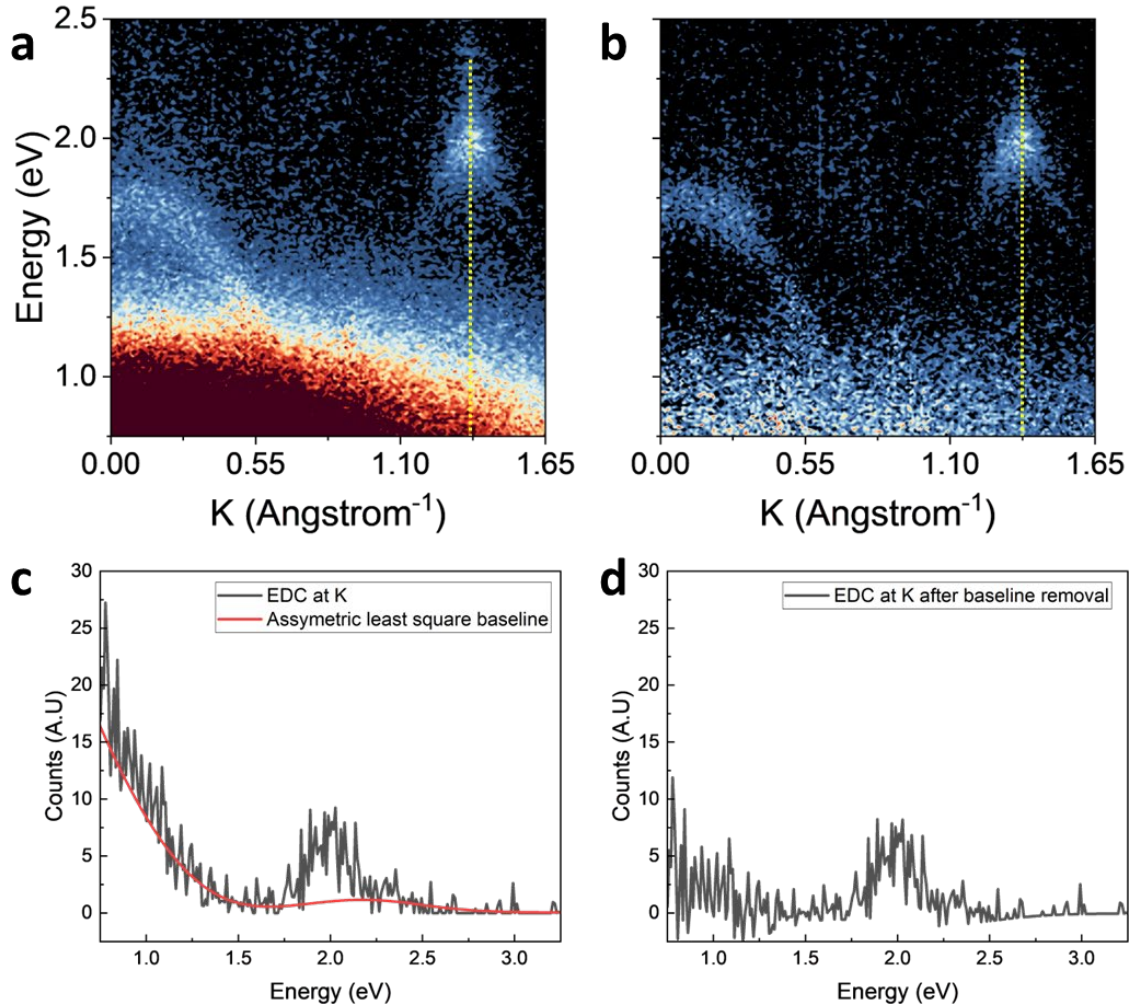

***SI figure 6. Background subtraction for Optically driven Floquet:** (a) Bandstructure showing the optically driven Floquet replica at 1.98 eV. The color profile is saturated to show the replica clearly. (b) The replica after the background subtraction. (c) EDC at the K point (yellow dotted line) along with the background obtained from the asymmetric least square smoothing algorithm. d) EDC at the same point (yellow dotted line) after the background subtraction. The Floquet replica is more clearly visible after the background subtraction.*

## Difference between Volkov and Floquet state

SI figure 7 shows the optical Floquet state and the Volkov state for 1.98 eV pump excitation. The Volkov and the Floquet state have different origins but have identical photoemission signals[42,49]. However, the two processes can be distinguished by choosing the correct pump polarization. For the *p*-polarized pump, the photoemission signal has contributions from both the Volkov and the Floquet states (SI figure 7b). For the case of *s*-polarization, only Floquet states contribute to the signal (SI figure 7a). This is further evidenced by the difference in the relative strengths of the replica signature we observe for the two different polarization cases. For the *p*- polarized excitation, the replica strength is 4% of the valence band, whereas the replica for the *s*-polarized case is only 2% of the valence band signal. The optically driven Floquet replica was observed for ~130 fs due to the finite pulse width of the pump and the probe pulse in our experiments (SI figure 7c). Moreover, the photoemission counts scale quadratically with the electric field strength as seen in the log-log plot in SI figure 7d.

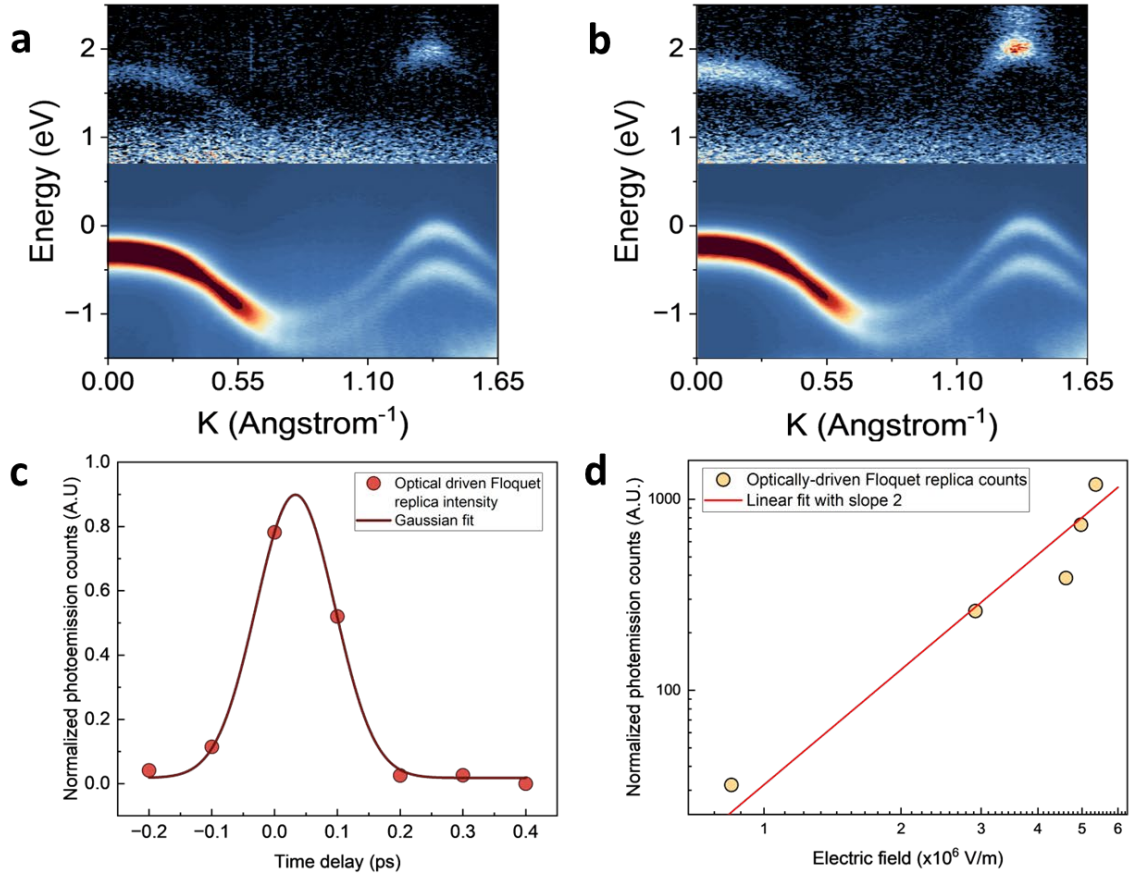

**SI figure 7. Comparison between Optically driven Floquet and pump-induced Volkov states:** *(a) Optically driven Floquet replica for the s-polarized pump pulse, (b) Pump-induced Volkov state together with optically driven Floquet replica for the p-polarized pump pulse. (c) The optically driven Floquet replica lasts for ~130 fs corresponding to the cross-correlation between the pump and the probe beams. The photoemission counts from the Optically driven Floquet replica was obtained by integrating a small momentum range (±0.2 Å⁻¹) and energy range (±0.2 eV) around the gamma point replica. The duration of the optically driven Floquet replica was obtained from the FWHM of the single gaussian fit to the photoemission counts (red curve). We expect the Volkov states to last for similar duration as optically driven Floquet replica. (d) log-log plot of the normalized photoemission counts from the optically driven Floquet replica versus the applied optical electric field strength. The photoemission counts show a quadratic behavior in applied electric field strength as confirmed by the slope 2 of the linear fit (red line) in the log-log plot.*

## Relation between TD-aGW self-energy, density, and exciton binding energy

In this section, we show that the effective electron-hole interaction is proportional to the square root of exciton density in the Method. We demonstrate this proportionality in a simple one-dimensional model, where the Hamiltonian is described by

$$H = \sum_{\mathbf{k}} (\epsilon_{v\mathbf{k}} a_{v\mathbf{k}}^{\dagger} a_{v\mathbf{k}} + \epsilon_{c\mathbf{k}} a_{c\mathbf{k}}^{\dagger} a_{c\mathbf{k}}) - U \sum_{\mathbf{k}} n_{\mathbf{k}} + \frac{U}{N} \sum_{\mathbf{k}_1, \mathbf{k}_2, \mathbf{q}} a_{v\mathbf{k}_1+\mathbf{q}}^{\dagger} a_{c\mathbf{k}_2-\mathbf{q}}^{\dagger} a_{c\mathbf{k}_2} a_{v\mathbf{k}_1},$$

where $\epsilon_{v\mathbf{k}}(\epsilon_{c\mathbf{k}})$ is the band energy of the valence (conduction) band with wavevector $\mathbf{k}$, U is the on-site interaction between two bands, and $N$ is the number of sites. The energy dispersion is given by $\epsilon_{v\mathbf{k}} = -2(1 - \cos \mathbf{k}) - E_g/2$, and $\epsilon_{c\mathbf{k}} = -\epsilon_{v\mathbf{k}}$ with $E_g = 1$ in our choice of system of units.

We demonstrate that the effective electron-hole interaction $\delta\Sigma_{vc\mathbf{k}}$ is proportional to the square root of carrier density in SI figure 8 by gathering data at different pump setup.

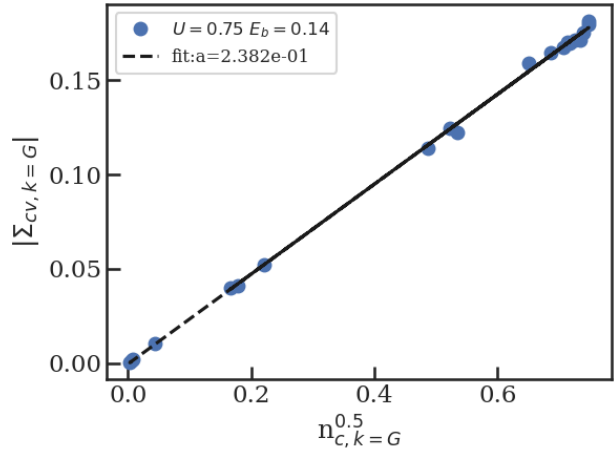

**SI figure 8.** *Effective interaction as a function of the square root of density.*